\documentclass[conference]{IEEEtran}
\usepackage{cite}

\ifCLASSINFOpdf
  \usepackage[pdftex]{graphicx}
  \usepackage{epstopdf}
  \DeclareGraphicsExtensions{.eps}
  \usepackage{float}
\else
  \usepackage[dvips]{graphicx}
  \DeclareGraphicsExtensions{.eps}
\fi
\usepackage{amsmath}
\usepackage{array}

\ifCLASSOPTIONcompsoc
  \usepackage[caption=false,font=normalsize,labelfont=sf,textfont=sf]{subfig}
\else
  \usepackage[caption=false,font=footnotesize]{subfig}
\fi
\usepackage{url}

\usepackage{multirow}

\usepackage{fixme}
\fxsetup{inline=true,margin=false}

\usepackage{hyperref}

\newcommand{\scriptV}{\mathcal{V}}
\newcommand{\scriptE}{\mathcal{E}}
\newcommand{\new}{\text{new}}
\newcommand{\prev}{\text{prev}}

\begin{document}
%
\title{Evaluating Link Prediction Accuracy on Dynamic Networks with Added 
and Removed Edges}

\author{\IEEEauthorblockN{Ruthwik R. Junuthula, Kevin S. Xu, and Vijay K. 
Devabhaktuni}
\IEEEauthorblockA{EECS Department, University of Toledo\\
2801 W. Bancroft St. MS 308, Toledo, OH 43606-3390, USA\\
\url{rjunuth@utoledo.edu}, 
\url{kevin.xu@utoledo.edu}, \url{vijay.devabhaktuni@utoledo.edu}}}


%


\maketitle

\begin{abstract}
The task of predicting future relationships in a social network, known as 
\emph{link prediction}, has been studied extensively in the literature. 
Many link prediction methods have been proposed, ranging from 
common neighbors to probabilistic models. 
Recent work by Yang et al.~\cite{Yang2014} has highlighted several challenges 
in evaluating link prediction accuracy. 
In dynamic networks where edges are both \emph{added and removed} over time, 
the link prediction problem is more complex and involves 
predicting both newly added and newly removed edges. 
This results in new challenges in the evaluation of dynamic 
link prediction methods, and the recommendations provided by 
Yang et al.~\cite{Yang2014} 
are no longer applicable, because they do not address edge removal. 
In this paper, we investigate several metrics currently used for evaluating 
accuracies of dynamic link prediction methods and demonstrate why they can be 
misleading in many cases. 
We provide several recommendations on evaluating dynamic link prediction 
accuracy, including separation into two categories of evaluation. 
Finally we propose a unified metric to characterize link prediction accuracy 
effectively using a single number.
\end{abstract}


%
\IEEEpeerreviewmaketitle

\section{Introduction}
The popularity of online social networking services has provided people with 
myriad new platforms for social interaction. 
Many social networking services also offer personalized suggestions of 
other people to follow or interact with, as well as websites or products 
that a user may be interested in. 
A key component in generating these personalized suggestions involves 
performing \emph{link prediction} on social networks.

The traditional problem of link prediction on networks is typically defined 
as follows: 
given a set of vertices or nodes $\scriptV$ and a set of edges or links 
$\scriptE$ connecting pairs of nodes, output a list of scores for all pairs 
of nodes without edges, i.e.~all pairs $(u,v) \notin \scriptE$, where a higher  
score for a pair $(u,v)$ denotes a higher predicted likelihood of an edge 
forming between nodes $u$ and $v$ at a future time\footnote{Link prediction 
is also used to predict missing edges in partially observed networks, 
where the score denotes the predicted likelihood of an unobserved edge 
between $u$ and $v$.}. 
Many link prediction methods have been proposed; see 
\cite{Liben-Nowell2007,AlHasan2011} for surveys of the literature. 
 
In this paper, we consider a more complex dynamic network 
setting where edges are both \emph{added 
and removed} over time, which is often referred to as 
\emph{dynamic link prediction} \cite{Xu2014a} or forecasting 
\cite{Foulds2011}. 
For instance, in a social network with timestamped edges denoting 
interactions between people (nodes), an edge may appear at several time 
instances where a pair of people are frequently interacting then disappear 
after interactions cease. 
Since existing edges between nodes may be removed at a future time, the 
dynamic link prediction problem is more complex and also involves computing 
a predicted score for \emph{existing} edges, because they may disappear at a 
future time. 

Evaluating link prediction accuracy involves comparing a binary 
label (whether or not an edge exists) with a real-valued predicted 
score. 
There are a variety of techniques for evaluation in this setting, including 
fixed-threshold methods such as F1-score and 
variable-threshold methods such as the area under the Receiver Operating 
Characteristic (ROC) curve, or AUC, and the area under 
the Precision-Recall (PR) curve, or PRAUC. 
Yang et al.~\cite{Yang2014} provide a comprehensive study of evaluation 
metrics for the traditional link prediction problem. 
Due to the severe class imbalance in link prediction (because only a small 
fraction of node pairs form edges), it was recommended to use  
PR curves and PRAUC for evaluating link predictors rather than ROC curves and 
AUC.

\begin{table}[t]
\caption{Illustration of disagreement among current metrics used to 
evaluate dynamic link prediction accuracy}
\label{tab:Disagree}

\centering
\renewcommand{\arraystretch}{1.2}
\begin{tabular}{ccccc}
	\hline
	Method                      & AUC         & PRAUC       & Max.~F1-score \\
	\hline
	TS-Adj \cite{Cortes2003}    & 0.780       & {\bf 0.239} & {\bf 0.371}   \\
	TS-AA \cite{gunecs2016link} & 0.777       & 0.065       & 0.144         \\
	TS-Katz \cite{Huang2009}    & {\bf 0.879} & 0.077       & 0.149         \\
	SBTM \cite{Xu2015}          & 0.799       & 0.138       & 0.337         \\
	\hline
\end{tabular}
\end{table}

To the best of our knowledge, there has not been prior work on evaluating 
accuracy in the dynamic link prediction or forecasting setting we 
consider. 
Prior studies on dynamic link prediction have typically used 
AUC \cite{Huang2009,Foulds2011,Kim2013,Xu2014a,gunecs2016link}, 
log-likelihood \cite{Foulds2011,Kim2013,Heaukulani2013}, 
and maximum F1-score \cite{Kim2013} as evaluation metrics. 

The evaluation of several dynamic link prediction methods using current 
metrics is shown in Table \ref{tab:Disagree}. 
(We discuss these methods in further detail in Section 
\ref{sec:PaperMethods}.) 
The table shows a clear \emph{disagreement} between current metrics for 
dynamic link prediction accuracy.
TS-Katz \cite{Huang2009} has the highest AUC but a low PRAUC and 
maximum F1-score, while TS-Adj \cite{Cortes2003} has highest PRAUC and 
maximum F1-score, but lower AUC. 
The SBTM \cite{Xu2015} ranks second in all three metrics. 
\emph{Which of these four methods is most accurate?} 
We seek to answer this question in this paper. 
This type of disagreement among evaluation metrics has also been observed in 
prior studies, including \cite{Kim2013}, but has not been investigated 
further. 

Inspired by the work of Yang et al.~\cite{Yang2014} in the traditional 
link prediction setting, we provide a thorough investigation of evaluation 
metrics for the dynamic link prediction problem. 
Our aim is \emph{not} to identify the most accurate 
link prediction algorithm, but rather 
to establish a set of recommendations for fair and effective evaluation of 
the accuracy of dynamic 
link prediction algorithms. 

Our main contributions are the following:
\begin{itemize}
\item We discuss why currently used metrics for dynamic 
link prediction can be misleading (Section \ref{sec:ExistingMetrics}).

\item We illustrate the importance of geodesic distance for the dynamic link 
prediction task and the dominance of edges at distance $1$ (Section 
\ref{sec:AnalysisofLinkProb}).

\item We separate the dynamic link prediction problem into two different link 
prediction problems based on geodesic distance 
and suggest metrics for fair and effective 
evaluation for each of the two problems (Section \ref{sec:TwoProblems}).

\item We propose a unified metric that characterizes link 
prediction accuracy using a single number and demonstrate that it avoids the 
shortcomings of currently used metrics for dynamic link prediction (Section 
\ref{sec:NewMetric}).
\end{itemize}

\section{Background}

\subsection{Problem Definition}
\label{sec:Setup}
The dynamic link prediction or forecasting problem is defined as follows. 
Given a set of nodes $\scriptV$ and a set of edges 
$\scriptE$ connecting pairs of nodes, output a list of scores for 
\emph{all pairs} 
of nodes, where a higher score for a pair $(u,v)$ denotes a higher 
predicted likelihood of an edge between $u$ and $v$ at a future time. 
Again, the main difference in the dynamic link prediction task 
compared to traditional link prediction is the need to output scores 
for node pairs where an edge is \emph{already present}, because the edge 
may be removed in the future. 

We consider dynamic networks observed at discrete time steps $1, 2, \ldots, T$. 
A common prediction setting used in time series forecasting is the rolling 
$1$-step forward prediction: for each $t=1, \ldots, T-1$, 
one trains a model using times $1$ to $t$ then predicts time $t+1$.  
In this paper, we perform dynamic link prediction in the rolling $1$-step 
forward prediction setting. 
The output of the link predictor contains $T-1$ sets of predicted scores 
for times $2$ to $T$ (trained using times $1$ to $T-1$, respectively),
which are then compared against $T-1$ sets of binary outputs denoting the 
actual states (edge or no edge) of all node pairs at times $2$ to $T$. 

To evaluate accuracy, we concatenate all of the predicted scores into a 
single vector and all of the binary outputs into a second vector. 
This setting has been adopted in many past studies including 
\cite{Foulds2011,Heaukulani2013,Kim2013,Xu2014a}. 
As noted in \cite{Yang2014}, we exclude node pairs corresponding to newly 
appearing nodes at any particular time step, since the identities of these 
new nodes are unknown at the time the prediction is computed. 

\subsection{Data Sets}

We use two data sets as running examples throughout this paper. 
The first is the NIPS co-authorship data collected by Globerson et 
al.~\cite{chechik2007eec}, consisting of papers
from the NIPS conferences from 1988 to 2003. 
Nodes in the NIPS data denote authors, and undirected edges denote 
collaborations between authors. 
Each year is used as a time step, and an edge between two nodes at a 
particular time step denotes that the authors co-wrote a paper 
together in the NIPS conference that year. 
The data set contains $2,865$ authors; we remove all authors who never 
collaborated with any other authors in the data set, leaving $2,715$ authors 
(nodes).

The second data set is the Facebook data collected by 
Viswanath et al.~\cite{Viswanath2009}. 
Nodes denote users, and directed edges represent interactions between 
users via posts from one user to another user's Facebook wall.
All interactions are timestamped, and we use $90$-day time steps (similar to the
analyses in \cite{Viswanath2009,Xu2015}) from the start of the data trace in June
2006, with the final complete $90$-day interval ending in November 2008, resulting
in $9$ total time steps. 
Viswanath et al.~collected data on over $60,000$ nodes. 
To make the dynamic link prediction problem more computationally tractable, 
we filter out nodes that have both in- and out-degree less than
$30$ in the aggregated network over all time steps, leaving $1,330$ nodes.
 
\begin{table}[t]
\renewcommand{\arraystretch}{1.2}
\centering
\caption{Summary statistics for data sets used in this paper. 
The last four rows show mean statistics over all time steps.}
\label{tab:DataSets}

\begin{tabular}{lcc}
	\hline
	& NIPS & Facebook \\
	\hline
	Directed           & No   & Yes      \\
	Number of time steps    & $17$ & $9$      \\
	Total number of nodes         & $2,715$      & $1,330$   \\
	\hline
	Mean number of edges         & $321$       & $3,714$          \\
	Mean edge probability       &  $1.7 \times 10^{-3}$      & $2.8 \times 10^{-3}$ \\
	Mean new edge probability   & $8.3 \times 10^{-5}$       & $1.4 \times 10^{-3}$          \\
	Mean prev.~observed edge probability & $0.031$               & $0.27$          \\
	\hline
\end{tabular}
\end{table}

Summary statistics for the two data sets are shown in Table 
\ref{tab:DataSets}. 
The edge probability at each time step is given by the number of actual edges 
divided by the number of possible edges, i.e.~the number of node pairs. 
We define a \emph{new edge} at time $t$ as an edge that did not appear in 
any time step $t' < t$. 
We define a \emph{previously observed edge} at time $t$ as an edge that 
appeared in at least one time step $t' < t$. 
Notice the large disparity between the new and previously observed edge 
probabilities---we will re-visit this point in Section \ref{sec:TwoProblems}. 

\subsection{Methods for Dynamic Link Prediction}
\label{sec:Methods}
Most methods for dynamic link prediction in the literature fall into one of 
three classes.

\subsubsection{Univariate Time Series Models}
Perhaps the most straightforward approach to dynamic link prediction is to 
apply standard univariate time series models to each node pair. 
Autoregressive Integrated Moving Average (ARIMA) models were used for  
dynamic link prediction in studies \cite{Huang2009,gunecs2016link}. 
A special case, the ARIMA($0,1,0$) model, 
is an exponentially-weighted moving average (EWMA) model, 
which has 
been used in studies \cite{Cortes2003,Dunlavy2011,Xu2011b,Xu2014a}.
Another approach is to model the probability of an edge between a pair of 
nodes to be proportional to the previous number of occurrences of that edge 
\cite{Dunlavy2011,Foulds2011,Kim2013,Heaukulani2013}, 
i.e.~a cumulative or growing window average, rather than an 
exponentially-weighted one. 
Dunlavy et al.~\cite{Dunlavy2011} referred to the EWMA as the collapsed 
weighted tensor and the cumulative average as the collapsed tensor.

Univariate time series approaches treat each node pair separately by ignoring 
the rest of the network altogether. 
In doing so, the predictors based on univariate time series 
models are limited in their predictive ability; 
for instance, they only predict future occurrences of previously 
observed edges and \emph{cannot predict new edges}. 
Thus these predictors are often used as baselines for comparison purposes. 
In many cases, however, these baselines have proven to be surprisingly 
competitive in accuracy as evaluated by existing metrics such as 
AUC \cite{Foulds2011, Xu2014a}, which can be quite deceiving as we discuss 
in Section \ref{sec:NewMetric}. 

\subsubsection{Similarity-Based Methods}
Node similarity-based methods have been among the earliest 
proposed methods for the traditional link prediction problem. 
These methods exploit the large number of triangles that are observed 
empirically in networks such as friendship networks to predict new edges. 
Typically used methods include common neighbors, Adamic-Adar, Jaccard 
coefficient, preferential attachment, and Katz \cite{Liben-Nowell2007}. 
These methods are often used in a static setting, where only a single snapshot 
of a network is available.

In the case of dynamic networks, these similarity-based methods have been 
used in several different manners. 
Huang and Lin \cite{Huang2009} aggregated the dynamic network over time to 
form a static network then apply similarity-based methods. 
G{\"u}ne{\c{s}} et al.~\cite{gunecs2016link} computed node similarities at each 
time step then model these similarities using ARIMA models. 
Dunlavy et al.~\cite{Dunlavy2011} proposed a truncated version of the Katz 
predictor based on a low-rank approximation of a weighted average of past 
adjacency matrices. 
From these studies, it appears that the Adamic-Adar and Katz predictors have 
been the most accurate among the similarity-based predictors.

Similarity-based methods have the opposite weakness of link predictors based 
on univariate time series models; that is, they ignore whether an edge has 
occurred in the past between a pair of nodes. 
These methods are sometimes used together with univariate time series models 
in practice \cite{Huang2009,gunecs2016link}.

\subsubsection{Probabilistic Generative Models}
\label{sec:Generative}
An alternative approach for dynamic link prediction is to fit a 
probabilistic generative 
model to the sequence of observed networks. 
A generative model for a dynamic network represents the network (up to time 
$t$) by a set of unobserved parameters $\Phi^t$. 
Given the values of the parameters, it then provides a model for the 
probability of an edge between any pair of nodes $(u,v)$ at time $t+1$, 
which is used as the link prediction score for $(u,v)$. 
Since the parameters $\Phi^t$ are unobserved, one 
typically estimates them from the sequence of networks then uses the 
estimated parameters to compute the link prediction score. 
The link prediction or forecasting accuracy is often used as a measure of 
goodness-of-fit for the generative model. 

Several classes of generative models for dynamic networks have been proposed, 
including dynamic latent feature models and dynamic stochastic block models. 
In a latent feature model, every node in a network has an unobserved 
(typically binary) feature vector. 
An edge between two nodes is then formed conditionally 
independently of all other node pairs given their feature vectors. 
These models have been adapted to dynamic networks by allowing the latent 
features to change over time \cite{Foulds2011,Heaukulani2013,Kim2013}. 
Such models have tremendous flexibility; however, 
fitting these models typically requires Markov chain Monte Carlo (MCMC) 
methods that scale up to only a few hundred nodes. 

Stochastic block models (SBMs) divide nodes into classes, 
where all nodes within a class 
are assumed to have identical statistical properties. 
An edge between two nodes is formed independently of all other node pairs 
with probability dependent only on the classes of the two nodes, giving the 
adjacency matrix a block structure where blocks correspond to pairs of 
classes. 
SBMs have been extended to the dynamic network setting by allowing the edge 
probabilities and class memberships to change over time 
\cite{Yang2011,Xu2014a,Xu2015}. 
The models proposed in \cite{Xu2014a,Xu2015} can be fit using an extended 
Kalman filter 
and local search procedure that scales to a few thousand nodes, an order of  
magnitude larger than methods for fitting dynamic latent feature models. 

\subsubsection{Other Methods}
Dunlavy et al.~\cite{Dunlavy2011} proposed to use matrix and tensor 
factorizations, namely truncated singular value decomposition (TSVD) and 
canonical decomposition/parallel factors (CANDECOMP/PARAFAC or CP) tensor 
models, respectively.
Tylenda et al.~\cite{Tylenda2009} proposed a ``time-aware'' version of a 
local probabilistic model based on the maximum-entropy principle. 
The approach involves weighted constraints based on the times at which 
edges occurred.

\subsubsection{Methods Considered in This Paper}
\label{sec:PaperMethods}

In this paper, we consider methods from each of the first three categories:
\begin{itemize}
\item TS-Adj \cite{Cortes2003}: a univariate time series model applied to each 
node pair.

\item TS-AA \cite{gunecs2016link}: a similarity-based method that extends the 
Adamic-Adar link predictor to the dynamic setting by applying a time series 
model to the Adamic-Adar scores over time for a node 
pair\footnote{Adamic-Adar is not applicable to directed networks so we first 
convert the Facebook network to an undirected network before applying TS-AA.}.

\item TS-Katz \cite{Huang2009}: a similarity-based method that extends the 
Katz predictor to the dynamic setting by applying a time series 
model to the Katz scores over time for a node pair\footnote{The 
approach is slightly different from what was proposed in \cite{Huang2009} 
and is similar to the approach used in \cite{gunecs2016link} for TS-AA; we 
find this approach to be almost universally more accurate than the 
approach in \cite{Huang2009}.}.

\item SBTM \cite{Xu2015}: a probabilistic generative model based on 
stochastic block models.
\end{itemize} 
We emphasize again that the objective of this paper is \emph{not} to identify 
the best prediction algorithm, thus this list is not exhaustive. 
For simplicity, we use the EWMA, which corresponds to ARIMA$(0,1,0)$ with 
forgetting or decay factor of $0.5$ as the 
time series model for each of the methods with prefix TS. 
Higher accuracy is likely attainable by better model selection for the ARIMA 
model parameters, but it is outside the scope of this paper.

\section{Existing Evaluation Metrics}
\label{sec:ExistingMetrics}
The currently employed evaluation methods discussed in the introduction and 
shown in Table \ref{tab:Disagree} indicate 
the lack of a principled metric, which makes it difficult to evaluate the 
accuracies of dynamic link prediction methods. 
Most of the evaluation metrics used in link prediction have been borrowed from
other applications such as information retrieval and classification.
Hence these metrics are naturally biased to favor certain aspects over others, 
which may result in either over- or under-representing the accuracy of a 
particular method.

\begin{table}[t]
	\centering
	\renewcommand{\arraystretch}{1.2}
	\caption{Confusion matrix for binary prediction}
	\label{tab:Confusion}
	\begin{tabular}{ccc}
		\hline
		& Predicted $1$'s ($p$) & Predicted $0$'s ($n$) \\ \hline
		
		Actual $1$'s ($P$) & True Positives ($TP$) & False Negatives ($FN$) \\
		
		Actual $0$'s ($N$) & False Positives ($FP$) & True Negatives ($TN$) \\
		\hline
	\end{tabular}
\end{table}

The output of a link predictor is usually a set of real-valued scores, 
which are compared against a set of binary labels, where each label denotes 
the presence ($1$) or absence ($0$) of an edge. 
One technique for comparison is to threshold the scores at a fixed 
value, transforming the real-valued scores into binary predictions. 
These binary predictions can then be compared against the binary labels 
by computing the confusion matrix shown in Table \ref{tab:Confusion} then 
using metrics based on the confusion matrix. 
A second technique involves sweeping the threshold over the entire range of 
predicted scores and plotting a threshold curve displaying the variation of 
one metric against another. 
A third technique, applicable only to probabilistic models, is to evaluate 
the likelihood of the model given the set of binary labels.

\subsection{Information Retrieval-Based Metrics}
\label{sec:IRMetrics}
In information retrieval, one is typically concerned with two metrics 
calculated from the confusion matrix in Table \ref{tab:Confusion}: 
precision (${\frac{TP}{TP+FP}}$) and recall (${\frac{TP}{TP+FN}}$). 
Precision and recall  
are often combined into a single measure using their harmonic mean, known 
as the F1-score ($ 2 \cdot \frac{\text{recall} \cdot \text{precision}} 
{\text{recall} + \text{precision}}$).

The precision, recall, and F1-score all vary with the choice of threshold 
applied to the real-valued scores. 
As an alternative to choosing a threshold, one sometimes computes the 
precision at $k$, also known as the top $k$ predictive rate, which denotes the 
number of correctly predicted links from the top $k$ scores. 
In the traditional link prediction setting, $k$ is typically chosen to be 
equal to 
the number of actual new edges $P$ \cite{Liben-Nowell2007}. 
Relative metrics are also used, such as the improvement in top $k$ predictive 
rate as compared to expected rate of a random predictor 
\cite{Liben-Nowell2007}. 
Yang et al.~\cite{Yang2014} discussed and empirically demonstrated several 
shortcomings of using fixed-threshold 
metrics in the traditional link prediction setting, 
which led to unstable results 
and disagreements as the threshold was varied.
We observe these shortcomings also in the dynamic link prediction setting. 

An alternative to fixed-threshold metrics is to use threshold curves, which 
work by shifting the threshold, computing the confusion matrix 
for each threshold, and finally computing metrics based on the confusion 
matrices.  
Threshold curves for different predictors are often compared using a single 
scalar measure, typically the area under the curve. 
In information retrieval, the commonly used threshold curve is the 
Precision-Recall (PR) curve. 
We denote the area under the PR curve by PRAUC. 
Simply linearly 
interpolating between points on the PR curve has been shown to be 
inappropriate for calculating PRAUC; we use the proper 
interpolation approach as discussed in \cite{Davis2006}.

PR curves consider only prediction of the positives and are generally used for 
needle-in-haystack problems common in information retrieval, where
negatives dominate and are not interesting.
For link prediction, PR curves give credit for correctly predicting edges but 
do not give credit for correctly predicting non-edges. 
Due to the sparsity of most types of networks including social networks, 
the number of non-edges is much greater than the number of edges, so 
Yang et al.~\cite{Yang2014} recommend the use of PRAUC for evaluation in 
the traditional link prediction setting. 

\subsubsection{Uses in Dynamic Link Prediction}
In the dynamic link prediction setting, 
Kim et al.~\cite{Kim2013} proposed to use the maximum F1-score over all 
possible threshold values, i.e.~identifying the point on the PR curve that 
maximizes F1-score. 
In this manner, it utilizes a single threshold that is determined by sweeping 
the PR curve rather than choosing a threshold a priori. 
This metric displays similar evaluation properties as PRAUC due to its 
dependence on the PR curve.

The normalized discounted cumulative gain (NDCG) over the 
top $k$ link prediction scores \cite{Tylenda2009} is another information 
retrieval-based metric that been used for evaluating dynamic link prediction 
accuracy. 
It is a fixed-threshold metric that suffers from the same drawbacks as 
other fixed-threshold metrics as discussed by Yang et al.~\cite{Yang2014}. 

\subsubsection{Shortcomings for Dynamic Link Prediction}
We argue that the PR curve is inappropriate 
for dynamic link prediction because it only considers the edges (positives). 
Accurate prediction of existing edges that \emph{do not appear} at a future 
time (negatives), is an important aspect of dynamic link prediction and 
\emph{is not captured by the PR curve}! 
Thus the PR curve and metrics derived from the PR curve, such as PRAUC 
and maximum F1-score, may be highly deceiving in the dynamic link prediction 
setting. 
Notice from Table \ref{tab:Disagree} that the most accurate link predictor 
according to PRAUC and maximum F1-score is the TS-Adj baseline predictor that 
\emph{does not predict any new edges}!
We expand on this discussion in Section \ref{sec:PrevLinks}.

\subsection{Classification-Based Metrics}
In classification, the commonly used metric is classification accuracy 
($\frac{TP+TN}{P+N}$ for binary classification) over all data points, 
which are node pairs in the case of link prediction. 
Classification accuracy is often deceiving in the case of highly imbalanced 
data, where high accuracy can be obtained even by a random predictor. 

In binary classification, one is often concerned with the 
true positive rate ($TPR = {\frac{TP}{TP+FN}}$) 
and false positive rate ($FPR = {\frac{FP}{FP+TN}}$), which can be 
calculated from the confusion matrix in Table \ref{tab:Confusion} for a 
fixed threshold. 
By sweeping the threshold, one arrives at the Receiver Operating 
Characteristic (ROC) curve. 
Different ROC curves are typically compared using the area under the ROC 
curve (AUC or AUROC). 

\subsubsection{Uses in Dynamic Link Prediction}
The AUC gives a single value that can be used to compare accuracy against
other models and is the most commonly used metric for evaluating dynamic 
link prediction accuracy 
\cite{Huang2009,Foulds2011,Dunlavy2011,Kim2013,Xu2014a,gunecs2016link}. 
The main difference compared to the traditional link prediction task is 
that the AUC is computed over all possible node pairs, not only node pairs 
without edges. 

G{\"u}ne{\c{s}} et al.~\cite{gunecs2016link} also evaluated AUC over smaller 
subsets of node pairs, such as node pairs with no edges over the past $3$ 
time steps.
Splitting up the evaluation into different subsets is a step in the right 
direction; however, G{\"u}ne{\c{s}} et al.~\cite{gunecs2016link} chose the 
subsets in a somewhat ad-hoc fashion and still rely on AUC over all node pairs 
as an evaluation metric, which is problematic as we discuss in the following. 
We present a principled approach for splitting up the evaluation of 
dynamic link prediction accuracy in Section \ref{sec:TwoProblems}. 

\subsubsection{Shortcomings for Dynamic Link Prediction}
Yang et al.~\cite{Yang2014} claimed that AUC is deceiving for evaluation 
of accuracy in the traditional link prediction setting due to the locality 
of edge formation.  
They found empirically that the probability of forming a new edge 
between a pair of nodes decreases as the geodesic (shortest path) distance 
between the node pair increases. 
We demonstrate in Section \ref{sec:AnalysisofLinkProb} that this problem is 
even greater in the dynamic link prediction setting, where edges at 
distance $1$, i.e.~edges that have been previously observed, are also 
considered in the evaluation. 

One of the appealing properties of AUC is its interpretation as the 
probability of a randomly selected positive 
instance appearing above a randomly selected negative instance 
\cite{Fawcett2006}. 
In the traditional classification setting, where instances are assumed to 
be independent and identically distributed (iid), this interpretation 
can be very useful. 
However, as we demonstrate in Section \ref{sec:AnalysisofLinkProb}, node 
pairs are certainly \emph{not} iid, and edge formation probabilities vary 
greatly based on whether an edge has previously existed. 
Using only this information, one can construct a predictor that achieves 
high AUC, as evidenced by the TS-Adj predictor in Table \ref{tab:Disagree}. 
Hence pooling together all node pairs to evaluate AUC can be highly 
deceiving. 

\subsection{Likelihood-Based Metrics}
Given a probabilistic model for observed data, 
the likelihood of a set of parameters is given by the probability of the 
observations given those parameter values. 
Since the actual parameter values are unknown, one typically calculates the 
likelihood using optimal parameter estimates or the estimated posterior 
distribution of the parameters given the observed data. 
It is often easier and more numerically stable to work with the log-likelihood 
rather than the likelihood 
itself, so the log-likelihood of a model is usually reported in practice.

\subsubsection{Uses in Dynamic Link Prediction}
Likelihood-based metrics are often used for evaluating link prediction 
accuracy for generative models and are a natural fit given their probabilistic 
nature.  
In the dynamic link prediction setting, 
the observations correspond to the observed network $1$ time step forward, 
i.e.~at time ${t+1}$, while the parameters correspond to the generative model  
parameters at time $t$ as discussed in Section \ref{sec:Generative}. 
The log-likelihood has been used in studies
\cite{Foulds2011,Heaukulani2013,Kim2013} as a metric for dynamic link 
prediction accuracy. 
Researchers often also calculate the log-likelihood of a baseline model, which 
is then used to measure relative improvement of a proposed model in terms of 
log-likelihood. 
For instance, studies \cite{Foulds2011,Heaukulani2013} use a Bayesian  
interpretation of a cumulative average as a baseline model. 

\subsubsection{Shortcomings for Dynamic Link Prediction}
In general, log-likelihoods may be very complex to calculate due to the 
effects of constant terms that are usually ignored when maximizing the 
log-likelihood. 
Additionally it is not possible
to obtain likelihood values for link predictors that are 
not based on probabilistic models.
Thus the scope of this metric is limited both by its complexity and 
applicability to only a small subset of link prediction techniques.

\section{The Effect of Geodesic Distance on Dynamic Link Prediction}
\label{sec:AnalysisofLinkProb}
One of the main differences between the typical machine learning setting and 
the link prediction setting is that node pairs are \emph{not} independent and 
identically distributed (iid). 
It has been shown that the probability of forming an edge between two nodes 
is highly dependent on the length of the shortest path between them, often 
called the \emph{geodesic distance} or just the distance. 
In the traditional link prediction problem, most edges are formed at geodesic 
distance $2$, and the probability of edge formation generally decreases 
monotonically with increasing geodesic distance \cite{Yang2014}. 

\begin{figure}[t]
	\centering
	\subfloat[]{\label{fig:LinkFormation}
	\includegraphics[width=1.65in]{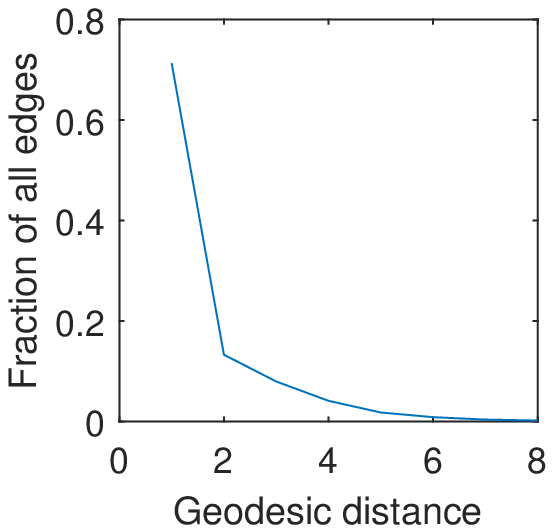}} \;
	\subfloat[]{\label{fig:DistProb}
	\includegraphics[width=1.65in]{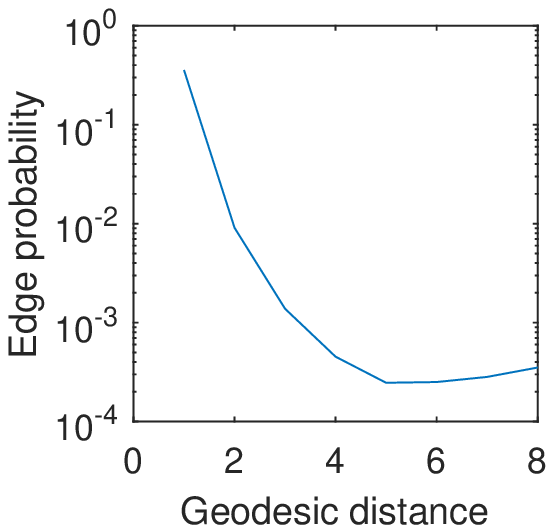}}
	\caption[]
	{\subref{fig:LinkFormation} 
	Fraction of all edges formed at each geodesic distance in the 
	Facebook data. 
	Each point denotes the number of edges formed at a certain 
	geodesic distance divided by the total number of edges formed at all 
	distances.
	\subref{fig:DistProb} 
	Empirical probability of forming an edge at 
	each geodesic distance in the Facebook data. 
	Each point denotes the number of edges formed at a certain geodesic 
	distance divided by the number of node pairs at that distance.}
	\label{fig:GeoDist}
\end{figure}

In the dynamic link prediction problem, we also need to consider 
node pairs at geodesic distance $1$, i.e.~pairs of nodes for which an edge 
has previously been formed, because these edges may or may not re-occur in 
the future. 
In the Facebook data set, we find that the majority (almost $80\%$) of edges 
are formed at distance $1$, as shown in Fig.~\ref{fig:LinkFormation}. 
Additionally Fig.~\ref{fig:DistProb} shows the empirical probability that 
an edge is formed between two nodes as a function of geodesic distance.
Notice that the edge probability is over $30$ times higher at distance 
$1$ compared to distance $2$ and over $300$ times higher than 
at distances $3$ and above! 
Thus it does not make sense to pool over all node pairs when evaluating 
dynamic link prediction accuracy (e.g.~using AUC or PRAUC), 
because the overwhelming majority of 
positive instances occur at distance $1$!

In the traditional link prediction problem, Yang et al.~\cite{Yang2014} 
suggested to evaluate link prediction accuracy separately at each distance. 
However this is a cumbersome approach, so they proposed also to use 
the PRAUC as a single measure of accuracy over all distances. 
As we have discussed in Section \ref{sec:ExistingMetrics}, PRAUC is 
problematic in the dynamic link prediction setting because it ignores the 
negative class, 
so we cannot use the same approach as in \cite{Yang2014}. 
Instead, recognizing that most edges are formed between node pairs with a 
previously observed edge, 
we propose to separate the dynamic link prediction problem into 
two problems.

\section{Separation into Two Link Prediction Problems}
\label{sec:TwoProblems}
Part of the difficulty in evaluating accuracy in the dynamic link prediction 
setting is related to the problem itself. 
Dynamic link prediction combines two problems: prediction of new links 
(distance $\geq 2$) and 
prediction of previously observed links (distance $= 1$). 
These two problems have very different properties in terms of difficulty, 
which primarily relates to the level of class imbalance in the two problems. 

The difference in difficulties of the two problems can be seen in Table 
\ref{tab:DataSets}. 
Notice that the probability of a \emph{new edge} being formed is tiny compared 
to a \emph{previously observed edge}! 
Thus the new link prediction problem involves much more severe class 
imbalance (i.e.~difficulty) compared to the previously observed link 
prediction problem. 
By pooling together all node pairs when calculating AUC or PRAUC, the 
evaluation is heavily biased towards the previously observed link prediction 
problem. 
As a result, all of the metrics shown in Table \ref{tab:Disagree} are 
biased in this manner. 
Instead, node pairs corresponding to possible new edges should be separated 
from node pairs corresponding to possible re-occurring edges, and accuracy 
metrics should be computed separately. 

\subsection{Prediction of New Edges}
\label{sec:NewLinks}

We begin by considering the prediction of new edges that have not been 
observed at any previous time. 
Actually this is simply the traditional link prediction problem, 
and the recommendations in \cite{Yang2014} apply here as well. 
The main recommendation is to use PR curves rather than ROC curves due to the 
abundance of true negatives as indicated by the extreme class imbalance shown 
in Table \ref{tab:DataSets}. 
By using PR curves, the overwhelming number of true negatives 
generated by link 
prediction algorithms are excluded from the evaluation. 

TS-Adj is capable only of predicting previously observed 
edges, as discussed in Section \ref{sec:Methods}. 
Thus its predictions for new links are random guesses, so it achieves the 
random baseline AUC of $0.5$ and PRAUC of $\frac{P}{P+N}$. 
The similarity-based methods TS-AA and TS-Katz are extensions of the 
Adamic-Adar and Katz predictors for traditional link prediction, and hence, 
they should be expected to perform better than the SBTM for new link 
prediction, especially because the SBTM does not consider geodesic distance. 
We see from Table \ref{tab:EvalSep} 
that this is indeed the case in both data sets, 
although the difference is much more pronounced in terms of PRAUC. 
Thus we support the recommendation in \cite{Yang2014} to use PRAUC to evaluate  
accuracy of new link prediction. 

\begin{table}[t]
	\renewcommand{\arraystretch}{1.2}
	\centering
	\captionsetup{position=top}
	\caption{Metrics for new and previously observed link 
	prediction}
	\label{tab:EvalSep}
	
	\subfloat[NIPS data]{\label{tab:EvalSepNips}
	\begin{tabular}{ccccc}
	\hline
	\multirow{2}{*}{Method} & \multicolumn{2}{c}{New Link} & \multicolumn{2}{c}{Prev.~Observed}      \\
	                            & AUC         & PRAUC $\times 10^{-3}$  & AUC         & PRAUC        \\
	\hline
	TS-Adj \cite{Cortes2003}    & 0.500       & 0.033                   & {\bf 0.855} & {\bf 0.099}  \\
	TS-AA \cite{gunecs2016link} & 0.534       & {\bf 0.882}             & 0.646       & 0.057        \\
	TS-Katz \cite{Huang2009}    & {\bf 0.535} & 0.735                   & 0.694       & 0.049        \\
	SBTM \cite{Xu2015}          & 0.531       & 0.055                   & 0.713       & 0.066        \\
	\hline
	\end{tabular}
	}
	\quad
	
	\subfloat[Facebook data]{\label{tab:EvalSepFacebook}
	\begin{tabular}{ccccc}
	\hline
	\multirow{2}{*}{Method} & \multicolumn{2}{c}{New Link} & \multicolumn{2}{c}{Prev.~Observed}     \\
	                            & AUC         & PRAUC $\times 10^{-3}$  & AUC         & PRAUC       \\
	\hline
	TS-Adj \cite{Cortes2003}    & 0.500       & 1.19                   & {\bf 0.705} & {\bf 0.417}  \\
	TS-AA \cite{gunecs2016link} & 0.712       & 14.4                   & 0.560       & 0.293        \\
	TS-Katz \cite{Huang2009}    & {\bf 0.768} & {\bf 14.8}             & 0.579       & 0.297        \\
	SBTM \cite{Xu2015}          & 0.700       & 4.41                   & 0.649       & 0.326        \\
	\hline
	\end{tabular}
	}
\end{table}

\subsection{Prediction of Previously Observed Edges}
\label{sec:PrevLinks}

The second problem in dynamic link prediction involves predicting edges 
that are currently present or were present at a previous time. 
As shown in Table \ref{tab:DataSets}, the class 
imbalance is several orders of magnitude \emph{less severe} than in the case 
of predicting new edges. 

Another major difference from new link prediction is the 
\emph{relevance of negatives (non-edges)}. 
Accurate prediction of negatives is \emph{highly relevant} 
because the removal of edges 
over time contributes a significant portion of the network dynamics. 
For example, in the NIPS co-authorship network, we find that over $85\%$ of 
edges observed at any time step are deleted at the following time 
step. 

A good evaluation metric for the task of predicting previously observed 
links must provide a balanced evaluation between the positive and negative 
classes. 
The metrics based on the PR curve are biased towards the positive class. 
We hence propose to use AUC, which is based on the ROC curve and \emph{does} 
account for negatives. 
Many of the shortcomings of AUC pointed out by Yang et al.~\cite{Yang2014} 
for the new link prediction task are not present in the previously 
observed link prediction task because the class imbalance is not nearly as 
significant. 

\begin{figure}[t]
\centering
\vspace{-10pt}
\subfloat[ROC curve]{\label{fig:ROC_Exist_AA_Katz}
\includegraphics[width=1.65in]{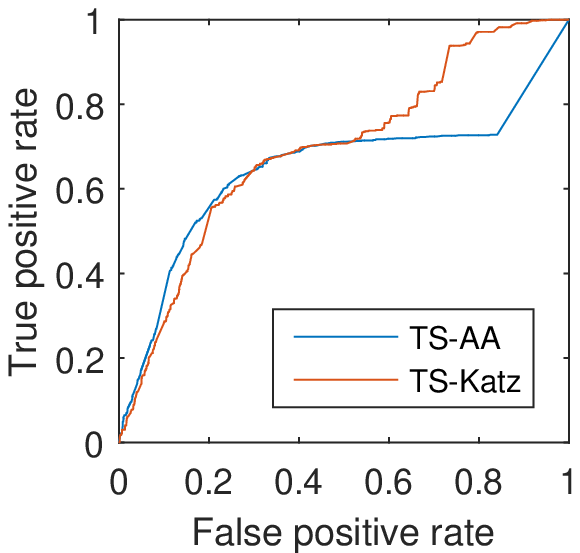}} \;
\subfloat[PR curve]{\label{fig:PR_Exist_AA_Katz}
\includegraphics[width=1.65in]{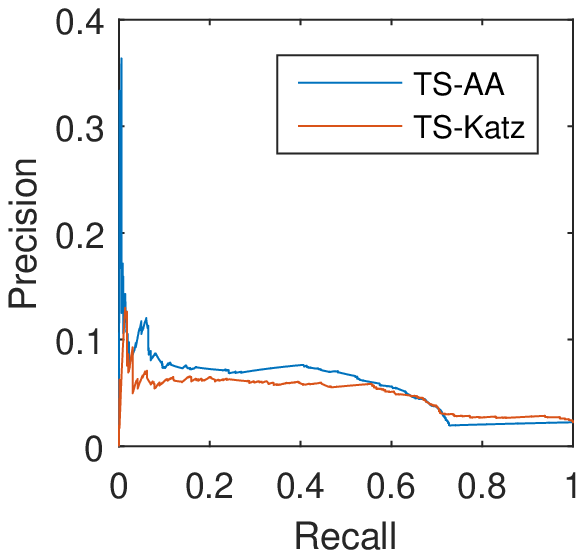}}
\caption[]
{Comparison of \subref{fig:ROC_Exist_AA_Katz} ROC and 
\subref{fig:PR_Exist_AA_Katz} PR curves for previously observed link 
prediction on NIPS data. 
TS-AA performs better at low recall (TPR) and worse at 
high recall, resulting in lower AUC but higher PRAUC.}
\label{fig:ROC_PR_Exist}
\end{figure}

\begin{figure}[t]
\centering
\subfloat[TS-AA]{\label{fig:AA_Scores}
\includegraphics[width=3in]{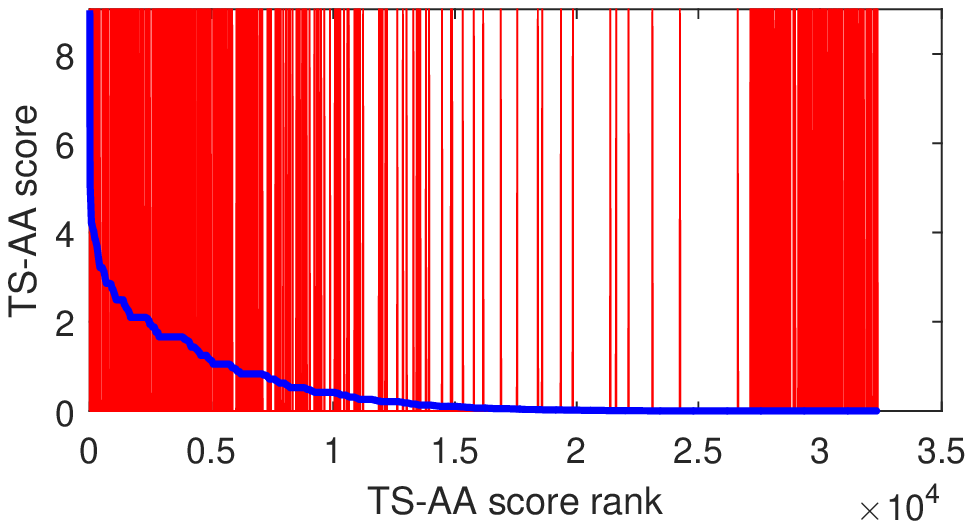}} \quad
\subfloat[TS-Katz]{\label{fig:Katz_Scores}
\includegraphics[width=3in]{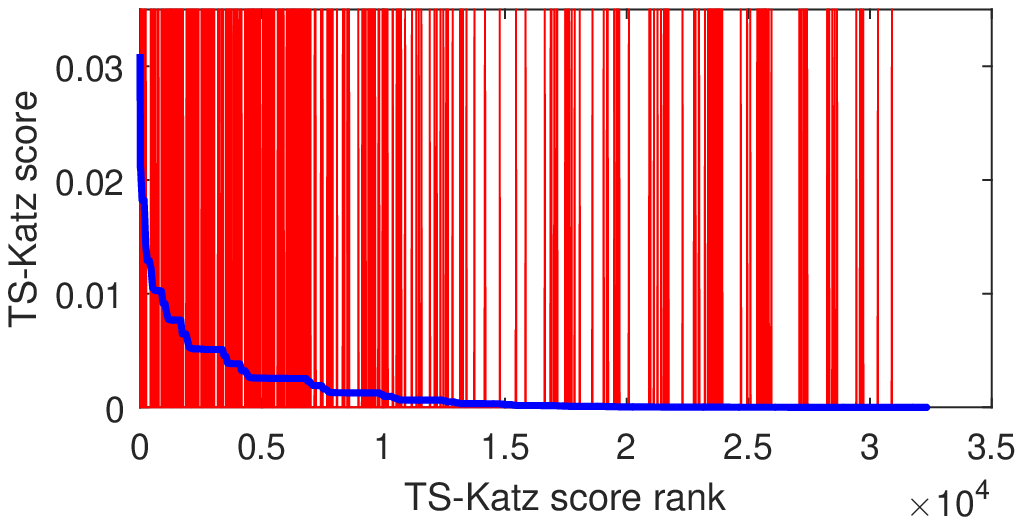}}
\caption[]
{Link prediction scores of \subref{fig:AA_Scores} TS-AA and 
\subref{fig:Katz_Scores} TS-Katz sorted in descending order (blue lines) 
corresponding to all node pairs for which an edge was previously observed. 
Red vertical lines denote node pairs that form an edge at the following 
time step. 
TS-AA correctly predicts more edges at high scores but misses many edges 
at low scores compared to TS-Katz.}
\label{fig:AA_Katz_Scores}
\end{figure}

From the AUC and PRAUC values for previously observed link prediction 
in Table \ref{tab:EvalSep},  
TS-Adj is the most accurate according to both metrics on both data sets. 
This is not surprising because TS-Adj can \emph{only} predict previously 
observed edges. 
However AUC and PRAUC do not necessarily agree in general; for example, 
consider TS-AA and TS-Katz on the NIPS data. 
TS-AA has higher PRAUC but lower AUC, and the reason for this can be seen 
in the ROC and PR curves shown in Fig.~\ref{fig:ROC_PR_Exist}. 
Fig.~\ref{fig:ROC_PR_Exist} can be further explained using 
Fig.~\ref{fig:AA_Katz_Scores}, where the sorted link 
prediction scores for TS-AA and TS-Katz are plotted with edges overlaid. 
TS-AA is more accurate than TS-Katz at high scores but worse at low scores, 
missing many edges. 
This leads to higher precision and lower FPR for low values of recall (TPR) but 
lower precision and higher FPR for high values of recall, which produces  
the disagreement between AUC and PRAUC.
Since the PR curve is only focused on accurate prediction of positives, TS-AA 
is rewarded for being more accurate at high scores (higher precision) and 
is less harshly penalized for missing edges at low scores. 
Thus we believe AUC to be a more balanced metric for evaluating accuracy of 
previously observed link prediction.

\section{A Unified Evaluation Metric}    
\label{sec:NewMetric}

By separating the dynamic link prediction problem into two problems 
with separate evaluation metrics, we are able to fairly evaluate different 
methods for dynamic link prediction. 
However one often desires a single metric to capture the ``overall'' 
accuracy rather than two metrics, analogous to the role of F1-score combining 
precision and recall. 

In the dynamic link prediction setting, any such metric should capture both 
the predictive power in new link and previously observed link prediction. 
In Section \ref{sec:TwoProblems}, we concluded that PRAUC is the better 
evaluation metric for new link prediction and that AUC is the better 
evaluation metric for previously observed link prediction. 
A unified evaluation metric could thus consist of the mean of the two 
quantities. 
Notice, however, from Table \ref{tab:EvalSep} that the two quantities 
have very large differences in 
magnitude, despite both being in the same range $[0,1]$. 
Thus the arithmetic mean is inappropriate because it would be dominated by 
the AUC value for previously observed link prediction. 
The harmonic mean 
is also inappropriate because it would be dominated by the PRAUC
for new link prediction, which has a much larger reciprocal.

We recommend instead to use the \emph{geometric mean} of the two quantities 
after a baseline correction, which we denote by
\begin{equation*}
\text{GMAUC} = \sqrt{\frac{\text{PRAUC}_{\new} - \frac{P}{P+N}} 
{1 - \frac{P}{P+N}} \cdot 2(\text{AUC}_{\prev} - 0.5)},
\end{equation*}
where $P$ and $N$ denote the number of actual edges and non-edges over the 
set of node pairs considered for new link prediction. 
The baseline correction subtracts the PRAUC and AUC values that would be 
obtained by a random predictor. 
The use of the geometric mean is motivated by the GMean metric proposed by 
Kubat et al.~\cite{kubat1997addressing} for evaluating classification accuracy 
in highly imbalanced data sets.
The geometric mean has several nice properties in this setting:
\begin{itemize}
\item It is based on threshold curves and avoids the pitfalls of 
      fixed-threshold metrics as discussed in \cite{Yang2014}.
      
\item It accounts for the different scales of the PRAUC for new edges and 
      AUC for previously observed edges without being dominated by either 
      quantity.
          
\item It is $0$ for any predictor that can only predict new 
      edges or can only predict previously observed edges.
\end{itemize}
The final point addresses an observation from several previous papers 
\cite{Foulds2011,Kim2013,Xu2014a} on generative models for dynamic networks: 
baseline methods (e.g.~TS-Adj) that predict only previous observed edges tend to perform 
quite competitively in terms of AUC when evaluated on the entire network. 
The GMAUC for a baseline predictor of this sort 
would be $0$ due to its inability to predict any new edges at all. 

\begin{table}[t]
	\renewcommand{\arraystretch}{1.2}
	\centering
	\captionsetup{position=top}
	\caption{Evaluation metrics for new and previously observed 
	link prediction and proposed GMAUC metric for unified evaluation}
	\label{tab:EvalProposed}
	
	\subfloat[NIPS data]{\label{tab:EvalProposedNips}
	\begin{tabular}{ccccc}
	\hline
	Method                      & PRAUC $\times 10^{-3}$ (new) & AUC (prev.) & GMAUC       \\
	\hline
	TS-Adj \cite{Cortes2003}    & 0.033                        & {\bf 0.855} & 0           \\
	TS-AA \cite{gunecs2016link} & {\bf 0.882}                  & 0.646       & 0.016       \\
	TS-Katz \cite{Huang2009}    & 0.735                        & 0.694       & {\bf 0.017} \\
	SBTM \cite{Xu2015}          & 0.055                        & 0.713       & 0.003       \\
	\hline
	\end{tabular}
	}
	\quad

	\subfloat[Facebook data]{
	\label{tab:EvalProposedFacebook}
	
	\begin{tabular}{ccccc}
	\hline
	Method                      & PRAUC $\times 10^{-3}$ (new) & AUC (prev.) & GMAUC       \\
	\hline
	TS-Adj \cite{Cortes2003}    & 1.19                         & {\bf 0.705} & 0           \\
	TS-AA \cite{gunecs2016link} & 14.4                         & 0.560       & 0.040       \\
	TS-Katz \cite{Huang2009}    & {\bf 14.8}                   & 0.579       & {\bf 0.047} \\
	SBTM \cite{Xu2015}          & 4.41                         & 0.649       & 0.031       \\
	\hline
	\end{tabular}
	}
\end{table}

The accuracies of several dynamic link predictors using the evaluation 
metrics proposed in this paper are shown in Table \ref{tab:EvalProposed}. 
According to the proposed GMAUC metric, TS-Katz is the best predictor for 
both data sets due to its ability to accurately 
predict both previously observed and new edges. 
Notice that, for the NIPS data, TS-Katz has the highest GMAUC despite not 
being the most accurate in either task. 
This is due to the balanced evaluation of new and previously observed 
link prediction used in the proposed GMAUC metric.

The data set used to compute the metrics shown in Table \ref{tab:Disagree} 
is actually the same Facebook data used in Table 
\ref{tab:EvalProposedFacebook}. 
Notice that the least accurate method according to all three metrics in Table 
\ref{tab:Disagree}, TS-AA, 
actually becomes the second most accurate once the evaluation is properly 
split up into new and previously observed links. 
This is primarily due to its strength in 
new link prediction compared to the SBTM, which does not consider geodesic 
distance for new link prediction, 
and to TS-Adj, which does not consider new edges at all. 

\section{Conclusions}
In this paper, we thoroughly examined the problem of evaluating accuracy in 
the dynamic link prediction setting where edges are both added and removed 
over time. 
We find that the overwhelming majority of edges formed at any given time are 
edges that have previously been observed. 
These edges should be evaluated \emph{separately} from new edges, i.e.~edges 
that have not formed in the past between a pair of nodes. 
The new and previously observed link prediction problems have very different 
levels of difficulty, with new link prediction being orders of magnitude more 
difficult. 
\emph{None of the currently used metrics for dynamic link prediction perform 
this separation} and are thus dominated by the accuracy on the easier problem 
of predicting previously observed edges.

Our main recommendations are as follows:
\begin{enumerate}
\item Separate node pairs for which edges have previously been observed from 
the remaining node pairs, and evaluate link prediction accuracy on these two 
sets separately.

\item For node pairs without previous edges, i.e.~the new link prediction 
problem, evaluate prediction accuracy using PRAUC due to the tremendous class 
imbalance.

\item For node pairs with previous edges, evaluate prediction accuracy using 
AUC due to the importance of predicting negatives (non-edges).

\item If a single metric of accuracy is desired, evaluate new and previously 
observed link prediction using separate metrics then combine the metrics 
rather than computing a single metric over all node pairs.

\item Use the proposed GMAUC metric as the single accuracy metric 
to provide a balanced evaluation between new and previously observed link 
prediction.  
\end{enumerate}


\bibliographystyle{IEEEtran}
\bibliography{SocialCom_2016_Evaluation}

\begin{thebibliography}{10}
\providecommand{\url}[1]{#1}
\csname url@samestyle\endcsname
\providecommand{\newblock}{\relax}
\providecommand{\bibinfo}[2]{#2}
\providecommand{\BIBentrySTDinterwordspacing}{\spaceskip=0pt\relax}
\providecommand{\BIBentryALTinterwordstretchfactor}{4}
\providecommand{\BIBentryALTinterwordspacing}{\spaceskip=\fontdimen2\font plus
\BIBentryALTinterwordstretchfactor\fontdimen3\font minus
  \fontdimen4\font\relax}
\providecommand{\BIBforeignlanguage}[2]{{%
\expandafter\ifx\csname l@#1\endcsname\relax
\typeout{** WARNING: IEEEtran.bst: No hyphenation pattern has been}%
\typeout{** loaded for the language `#1'. Using the pattern for}%
\typeout{** the default language instead.}%
\else
\language=\csname l@#1\endcsname
\fi
#2}}
\providecommand{\BIBdecl}{\relax}
\BIBdecl

\bibitem{Yang2014}
Y.~Yang, R.~N. Lichtenwalter, and N.~V. Chawla, ``Evaluating link prediction
  methods,'' \emph{Knowl.~Inform.~Sys.}, vol.~45, no.~3, pp. 751--782, 2015.

\bibitem{Liben-Nowell2007}
D.~Liben-Nowell and J.~Kleinberg, ``{The link-prediction problem for social
  networks},'' \emph{J.~Am.~Soc.~Inform.~Sci.~Tech.}, vol.~58, no.~7, pp.
  1019--1031, 2007.

\bibitem{AlHasan2011}
M.~{Al Hasan} and M.~J. Zaki, ``{A survey of link prediction in social
  networks},'' in \emph{Social network data analytics}.\hskip 1em plus 0.5em
  minus 0.4em\relax Springer, 2011, pp. 243--275.

\bibitem{Xu2014a}
K.~S. Xu and A.~O. {Hero III}, ``{Dynamic stochastic blockmodels for
  time-evolving social networks},'' \emph{IEEE J.~Sel.~Top.~Signal Process.},
  vol.~8, no.~4, pp. 552--562, 2014.

\bibitem{Foulds2011}
J.~R. Foulds, C.~DuBois, A.~U. Asuncion, C.~T. Butts, and P.~Smyth, ``{A
  dynamic relational infinite feature model for longitudinal social
  networks},'' in \emph{Proc.~14th Int.~Conf.~Artif.~Intell.~Stat.}, 2011, pp.
  287--295.

\bibitem{Cortes2003}
C.~Cortes, D.~Pregibon, and C.~Volinsky, ``{Computational methods for dynamic
  graphs},'' \emph{J.~Comput.~Graph.~Stat.}, vol.~12, no.~4, pp. 950--970,
  2003.

\bibitem{gunecs2016link}
{\.I}.~G{\"u}ne{\c{s}}, {\c{S}}.~G{\"u}nd{\"u}z-{\"O}{\u{g}}{\"u}d{\"u}c{\"u},
  and Z.~{\c{C}}ataltepe, ``Link prediction using time series of
  neighborhood-based node similarity scores,'' \emph{Data Min.~Knowl.~Discov.},
  vol.~30, no.~1, pp. 147--180, 2016.

\bibitem{Huang2009}
Z.~Huang and D.~K.~J. Lin, ``{The time-series link prediction problem with
  applications in communication surveillance},'' \emph{INFORMS J.~Comput.},
  vol.~21, no.~2, pp. 286--303, 2009.

\bibitem{Xu2015}
K.~S. Xu, ``{Stochastic block transition models for dynamic networks},'' in
  \emph{Proc.~18th Int.~Conf.~Artif.~Intell.~Stat.}, 2015, pp. 1079--1087.

\bibitem{Kim2013}
M.~Kim and J.~Leskovec, ``{Nonparametric multi-group membership model for
  dynamic networks},'' in \emph{Adv.~Neural Inform.~Process.~Sys.~25}, 2013,
  pp. 1385--1393.

\bibitem{Heaukulani2013}
C.~Heaukulani and Z.~Ghahramani, ``{Dynamic probabilistic models for latent
  feature propagation in social networks},'' in \emph{Proc.~30th
  Int.~Conf.~Mach.~Learn.}, vol.~28, 2013, pp. 275--283.

\bibitem{chechik2007eec}
A.~Globerson, G.~Chechik, F.~Pereira, and N.~Tishby, ``{Euclidean embedding of
  co-occurrence data},'' \emph{J.~Mach.~Learn.~Res.}, vol.~8, pp. 2265--2295,
  2007.

\bibitem{Viswanath2009}
B.~Viswanath, A.~Mislove, M.~Cha, and K.~P. Gummadi, ``On the evolution of user
  interaction in facebook,'' in \emph{Proc.~2nd ACM Workshop Online
  Soc.~Netw.}, 2009, pp. 37--42.

\bibitem{Dunlavy2011}
D.~M. Dunlavy, T.~G. Kolda, and E.~Acar, ``{Temporal link prediction using
  matrix and tensor factorizations},'' \emph{ACM Trans.~Knowl.~Discov.~Data},
  vol.~5, no.~2, p.~10, 2011.

\bibitem{Xu2011b}
K.~S. Xu, M.~Kliger, and A.~O. {Hero III}, ``{A shrinkage approach to tracking
  dynamic networks},'' in \emph{Proc.~IEEE Stat.~Signal Process.~Workshop},
  2011, pp. 517--520.

\bibitem{Yang2011}
T.~Yang, Y.~Chi, S.~Zhu, Y.~Gong, and R.~Jin, ``{Detecting communities and
  their evolutions in dynamic social networks---a Bayesian approach},''
  \emph{Mach.~Learn.}, vol.~82, no.~2, pp. 157--189, 2011.

\bibitem{Tylenda2009}
T.~Tylenda, R.~Angelova, and S.~Bedathur, ``{Towards time-aware link prediction
  in evolving social networks},'' in \emph{Proc.~3rd Workshop
  Soc.~Netw.~Min.~Anal.}, 2009, p.~9.

\bibitem{Davis2006}
J.~Davis and M.~Goadrich, ``{The relationship between Precision-Recall and ROC
  curves},'' in \emph{Proc.~23rd Int.~Conf.~Mach.~Learn.}, 2006, pp. 233--240.

\bibitem{Fawcett2006}
T.~Fawcett, ``{An introduction to ROC analysis},'' \emph{Pattern Recog.~Lett.},
  vol.~27, no.~8, pp. 861--874, 2006.

\bibitem{kubat1997addressing}
M.~Kubat and S.~Matwin, ``Addressing the curse of imbalanced training sets:
  one-sided selection,'' in \emph{Proc.~14th Int.~Conf.~Mach.~Learn.}, 1997,
  pp. 179--186.

\end{thebibliography}
%
%
%

\end{document}